\documentclass[referee]{aa}
\usepackage{graphicx}
\usepackage{txfonts}

\begin{document}

\title{Polarimetry of an Intermediate-age Open Cluster: NGC 5617 \thanks{Based on observations obtained at Complejo Astron\'omico El Leoncito, operated under agreement between the Consejo Nacional de Investigaciones Cient\'{\i}ficas y T\'ecnicas de la Rep\'ublica Argentina and the Universities of La Plata, C\'ordoba, and San Juan.}}

\author{Ana Mar\'{\i}a Orsatti
\inst{1}\fnmsep\inst{2}\fnmsep\inst{3}
        \and
  Carlos Feinstein
        \inst{1}\fnmsep\inst{2}\fnmsep\inst{3}
\and
M. Marcela Vergne
        \inst{1}\fnmsep\inst{2}\fnmsep\inst{3}
\and
Ruben E. Mart\'{\i}nez
\inst{1}\fnmsep\inst{2}
\and
E. Irene Vega
\inst{1}\fnmsep\inst{3}
        }

\institute{Facultad de Ciencias Astron\'omicas y Geof\'{\i}sicas, Observatorio Astron\'omico, Paseo del Bosque, 1900 La Plata, Argentina
\and Instituto de Astrof\'{\i}sica de La Plata, CONICET
\and Member of the Carrera del Investigador Cient\'{\i}fico, CONICET, Argentina}

\abstract{}{We present polarimetric observations in the UBVRI bands of 72 stars located in the direction of the medium age open cluster NGC 5617. Our intention is to use polarimetry as a tool membership identification, by building on previous investigations intended mainly to determine the cluster's general characteristics rather than provide membership suitable for studies such as stellar content and metallicity, as well as study the characteristics of the dust lying between the Sun and the cluster.}
{The obsevations were carried out using the five-channel photopolarimeter
of the Torino Astronomical Observatory attached to the 2.15m telescope at the Complejo Astron\'omico El Leoncito (CASLEO; Argentina).}
{We are able to add 32 stars to the list of members of NGC 5617, and review the situation for others listed in the literature.
In particular, we find that five blue straggler stars in the region of the cluster are located behind the same dust as the member stars are and we  confirm the membership of two red giants. The proposed polarimetric memberships are compared with those derived by photometric and kinematical methods, with excellent results.
Among the observed stars, we identify 10 with intrinsic polarization in their light.
NGC 5617 can be polarimetrically characterized with $ P_{max}= 4.40 \%$ and $ \theta_{v}= 73^\circ.1$. The spread in polarization values for the stars observed in the direction of
the cluster seems to be caused by  the uneven distribution of dust in front of the cluster's face.
Finally, we find that in the direction of the cluster, the interstellar medium is apparently free of dust, from the Sun's position up to the Carina-Sagittarius arm, where NGC 5617 seems to be located at its farthest border.} {}


\titlerunning{Polarimetry of NGC 5617}

\authorrunning{Orsatti et al.}\maketitle

\section{Introduction}

The polarimetric technique is a very useful tool for obtaining significant information (e.g. magnetic field direction,\ $ \lambda_{max}$, $P_{\lambda_{max}}$)
about  the dust located in front of a luminous object. Open clusters are very good candidates for carrying out polarimetric
observations, because previous photometric and spectroscopic studies of these clusters have provided detailed information about the
color and luminosity of the main-sequence stars.
In addition to the cluster physical parameters obtained by using those tools (e.g. age, distance, extinction, membership), the polarimetric data allow us study the location, size, and efficiency of the dust grains to polarize the starlight and the different directions of the Galactic magnetic field along the line of sight to the cluster. Since the open clusters are also spread within a fixed area, we can analyze the evolution in the physical parameters of the dust all over the region. In this framework, we  conduct systematic polarimetric observations in a large number of Galactic open clusters.

As part of this survey, we present a polarimetric investigation of the open cluster NGC 5617 (C1426-605). It is located at (l = $ 314\fdg7$, b = $ -0\fdg1$), covering a wide area of the sky of about 10 x 10 arc minutes. In the past it was investigated  using photoelectric and photographic photometry (Lindoff 1968; Moffat $ \&$ Vogt 1975; Haug 1978).
 Kjeldsen $ \&$ Frandsen (\cite{KF91}) and Carraro $ \&$ Munari (\cite{CM04}) presented CCD observations covering part of the area. In this last work deep CCD (BVI) photometry of the core region was performed to derive
more accurate estimates of the cluster fundamental parameters by observing  about 140 stars down to V = 17.5 mag.
The analysis of two adjacent fields covering the central part of the cluster confirmed a previous  mean value  of the
excess  $ E_{B-V}$= 0.48 mag, found in the first of  two CCD investigations, and a distance determination that located the open cluster at 2.0 $\pm$ 0.3 kpc from the Sun. It is an intermediate-age open cluster ($8.2~ 10^7$~years) containing red giants and blue straggler stars (Ahumada  \& Lapasset \cite{AL07}) in its surroundings, whose memberships of the cluster remain in doubt.

\section{Observations}
                                                                                
Observations of the $ UBV(RI)_{\rm KC}$ bands (KC: Kron-Cousins, $ \lambda_{U_\mathrm{eff}}$ =
0.36 $ \mu$m, FWHM = 0.05 $ \mu$m; $ \lambda_{B_\mathrm{eff}}$ = 0.44 $ \mu$m,
FWHM = 0.06 $ \mu$m; $ \lambda_{V_\mathrm{eff}}$ = 0.53 $ \mu$m, FWHM = 0.06 $ \mu$m; $ \lambda_{R_\mathrm{eff}}$ = 0.69 $ \mu$m,
FWHM = 0.18 $ \mu$m; $ \lambda_{I_\mathrm{eff}}$ = 0.83 $ \mu$m, FWHM = 0.15 $ \mu$m) were carried out using the five-channel photopolarimeter
of the Torino Astronomical Observatory attached to the 2.15m telescope at the Complejo Astron\'omico El Leoncito (CASLEO, Argentina). They were
performed over eight nights (June 27-30, 2006 and June 15-18, 2007). The instrument allows simultaneous polarization measurements to be performed in the five bands UBVRI, and the combination of the dichroic beam splitters and the filters cemented to the field lenses of the dichroic filters set closely matches the standard UBVRI system.
The FOTOR polarimeter provides the possibility of using 5 different diaphragm apertures to measure the stars, of sizes between 19.2 arcsec  and  5.6  arcsec. As usual, the same diaphragm was used to measure the star and the closest portion of sky.
During each observing run, a set of standard stars to determine null polarization and  the zero point of the polarization position angle (taken from Clocchiatti \& Marraco (\cite{CM88})) was observed to determine both the instrumental polarization and the coordinate transformation into the equatorial system, respectively. The net polarization of telescope and instrument, typically of about 0.01 percent, was subtracted from all the data.                                      
For additional information about the instrument, data acquisition, and data reduction, we refer to Scaltriti et al. (\cite{SPCACetal89}) and Scaltriti (\cite{S94}).

Table 1 lists the 72 stars observed polarimetrically in the direction of the open cluster, the percentage polarization
($ P_{ \lambda}$), the position angle of the electric vector ($ \theta_{\lambda}$) in the equatorial coordinate system,
and their respective mean errors for each filter. We also indicate the number of 60 second integrations for each filter. Star identifications are taken from Haug (1978).

\section{Results}

The sky projection of the V-band polarization vectors for the observed stars
in NGC 5617 are shown in Fig. 1. The cluster spreads over a large region covering more than 10 x 10 arcminutes.
The dotted line superimposed on the figure is the Galactic parallel of b=-0$\fdg$.2 In this plot we observe
that most of the stars  have their polarimetric vectors orientated in a direction close to the projection of the Galactic Plane (hereafter GP) in that region, but a close inspection of the angle distribution displays a complex structure.

Figure 2 (upper plot) shows the relation that exists between
$ P_{V}$ and $ \theta_{V}$. We can see a core of stars at $ P_V\sim$ 4\%  and  $ \theta_V\sim$76$^{o}$, and some scatter
toward lower angles and polarizations. This could be the result of a patchy dust distribution that produces  variable extinction, or the contamination by
non-member stars with polarizations originated in dust clouds other than those affecting the light from NGC 5617.
Unfortunately, papers based on CCD data did not perform a membership analysis of individual stars in the region so as to help disclose this point.
What can clearly be seen in this figure is that the orientation of the Galactic plane is similar to the orientation of the low polarization stars ($\sim$2\%), but this is not so for the stars in that core ($\sim$ 4\%).
This behavior is more clearly displayed by the histogram of the observed angles (lower plot). There is a peak around
75$^{o}$, but also a large scatter toward lower angles and a moderate scatter to the opposite side.
The dashed line is a Gaussian fit to the data, which provides a poor fit. The data for the peak region exhibit a
narrow and concentrated distribution, but a component that is broader than the best-fit Gaussian of the data.
This can be easily explained if each component, the concentrated peak and the more widely spread data, have different origins. The narrow component has the empirical regular shape found in other clusters (e.g., in NGC 6193, NGC 6167, NGC 6204, Hogg 22, Stock 16, Trumpler 27; from Waldhausen et al. \cite{WMF99}, Mart\'{\i}nez  et al. \cite{MVF04}, Feinstein et al. \cite{FBVNC00}, and Feinstein et al. \cite{FBVV03}, respectively) where the FWHM has values between 8 and 16 degrees, which seem to be compatible with the core width.

We understand that the most likely explanation of this behavior is that  the sample is contaminated by non-member stars of lower polarization and angle that are similar stars to those of the GP.  Although it is a statistically marginal result, 
the lower angle stars appear to be more numerous when the projected distance from the cluster's center increases. Carraro \& Munari  (\cite{CM04}) support this idea by showing that if they consider only stars inside a radius smaller than or equal to 2'.2, a clean main sequence is obtained, since most of the non-member stars are eliminated from their sample.

\section{Analysis and Discussion}

\subsection{Fitting with Serkowski's law}

To analyze the data, the polarimetric observations in the five filters were fitted for each star using Serkowski's law of interstellar
polarization (Serkowski 1973) given by  $$ P_{\lambda}/P_{\lambda max}=e^{-Kln^2(\lambda_{max}/\lambda)}. \  \  \ \ (1)$$
If polarization is produced by aligned interstellar dust particles, then we assume that, the observed data (in terms of
wavelength within the bands UBVRI) can be reproduced accurately by Eq (1) and that each star has a separate  $\lambda_{max}$ and a $P_{\lambda_{max}}$ value.

To perform the fitting, we adopted $ K=~1.66 \lambda_{max} + 0.01$, where $\lambda_{max}$ is in  $ \mu$m
(Whittet et al. \cite{WMHRetal92}). For each star,
we also computed the $\sigma_{1}$ parameter (the unit weight error of the fit) in order to quantify the departure of our data from the ``theoretical curve'' of the Serkowski's law. In our scheme, when a star exhibits $ \sigma_{1} > 1.80$ this is indicative of a non-interstellar origin (that is, an intrinsic polarization) in part of the measured polarization. The dominant source of intrinsic polarization is dust non-spherically distributed and, for classical Be stars, electron scattering.
The $\lambda_{max}$ values can also be used to test the origin of the polarization:  objects with a $\lambda_{max}$ much shorter than the
average value for the interstellar medium (0.55 $\mu m$, Serkowski et al. \cite{SMF75}) are also likely
to contain an intrinsic component of polarization (Orsatti et al. \cite{OVM98}).
The individual $P_{\lambda_{max}}$, $\sigma_{1}$, $ \lambda_{max}$, and $\bar{\epsilon}$ values, together with the star identification
from Haug (\cite{H78}), are listed in Table 2. We excluded five stars with $ \epsilon _{P_{max}}$ higher than 15$ \%$:  \#58, 69, 137, 149, and 255.
The mathematical expression used to obtain the individual $\sigma_{1}$ values is found in this table as a footnote.

According to Table 2, only 10  of the 67 stars exhibit signatures of intrinsic polarization: \#154,~260,~270, and 274 (a group with very high $\sigma_{1}$); \#146,~330, and 333 (with lower values): and also two blue stragglers (\#195,\#261) and the red giant \#227.
Star 226 has  $\sigma_{1}$= 1.91 but this value was estimated using data for only 3 filters, so the detection of intrinsic polarization in the star is dubious.
The use of the second criterion to detect intrinsic stellar polarization did not provide new candidates. 
Figure 3 shows, for some of these stars, both the polarization and position angle dependence on wavelength.
For comparison purposes, the best fit Serkowski's law for an interstellar origin of the polarization has been plotted as a
continuous line. In the individual plots, the ${P_\lambda} $ values do not appear to fit this law and in some cases (e.g., \#227,\#261) there is evidence of a combination of two different polarization mechanisms.
Most of the stars in the figure also show important rotations of the polarization position angle with $ \lambda$.

\subsection{The $ Q_{v}\  versus \  U_{v}$ plot and membership review}

The identification of members in a cluster is  important, not only to distance and age determinations but also 
other studies such as those of stellar content and metallicity. In NGC 5617, as in many other clusters, probable cluster members were identified as those stars simultaneously having reconcilable positions in both color-color and color-magnitude diagrams. 

Several errors in membership assignment are possible using this kind of photometric approach, for example when dealing with photographic photometry, in particular the U-measures; and when studying intermediate-age clusters, where the evolved stars could not be located close to the ZAMS. In our cluster, the CCD plates of Kjeldsen $ \&$ Frandsen (\cite{KF91}) and  Carraro $ \&$ Munari (\cite{CM04}) cover only the central region of the cluster, and for stars in the vicinities of the core only photographic measurements are possible. In addition, as found by these last authors, stars brighter than 12.5 mag are evolving away from the main sequence. Evolved members and background stars become mixed close to the ZAMS in the color-magnitude diagrams, and in this case the photometric identification of cluster members becomes difficult.

The polarimetric technique can help us to solve membership problems. Different plots used in combination with photometric information can be useful for separating between members and non-members. One of those plots presents the Stokes parameters $ Q_{v}\ versus\ U_{v}$ for the V-bandpass, where $ Q_{v} = P_{v}~ cos(2\theta_{v})$ and
{\bf $ U_{v} = P_{v}~ sin (2\theta_{v})$} are the components in the equatorial system of the
polarization vector $ P_{v }$,  is shown in Figure 4. The plot illustrates the variations occurring in interstellar environments. Since the light from cluster members must have traversed a common sheet of dust, of particular polarimetric characteristics, the member data points should occupy similar regions of the figure. Non-member stars (frontside and background stars) should be located in the $ Q_{v}\  versus \  U_{v}$ figure somewhat apart from the region occupied by member stars, since their light must have traveled through different dust clouds from those affecting the light of member stars, of different polarimetric characteristics.

The basic principle behind the use of polarimetry as a criterion for distinguishing members from non-members in a cluster is similar to that used in photometry to decide membership.  The procedure is based on the assumption that member stars are located behind common dust clouds that polarize their light, while this is not valid for most non-member stars.

Objects closer to the Sun, or located along the line of sight to the cluster, will have lower $ E_{B-V}$ and their light will be less polarized. If there are clouds between these stars and ourselves, and dust is orientated in a different direction, the final angle will not be the same as for the cluster's stars. In the Q versus U plot, these stars therefore are located in different regions.

For stars located behind the cluster, the individual polarizations could be higher than those associated with the cluster if the dust has the same orientation, or it could be depolarized if the orientation is not the same. However, in both cases the location in the diagram Q-U will not be the same as that for the cluster's stars.  And in the last case, the $ E_{B-V}$ of those stars will be higher and detectable in the efficiency diagram ( $ P_{v} $ versus $ E_{B-V}$).

For example,  the polarimetry became a very useful membership identification tool for stars belonging to the clusters NGC 6204 and Hogg 22, the last one located behind the first cluster.

Stars of each object occupy  different regions of the Q versus U diagram (see Fig. 6 of Mart\'{\i}nez et al. 2003), and since Hogg 22 is depolarized and has a higher $ E_{B-V}$'s, both clusters are separated in the polarization efficiency diagram (Fig. 5 of that work).

In using polarimetry  to decide memberships, we are able to use information  in the literature derive using alternative methods (e.g., photometric, spectroscopic, proper motions) and we can determine in the Q - U and $ P_{v}$ versus  $ E_{B-V}$ plots the regions where cluster and non-member stars are located. Because of previous considerations, we accept at first that any star in the non-member region is a very probable non-member and in the same way, any star located in the cluster region is also a very probable member of the cluster. We note  that we do not apply this criterion to stars with detected intrinsic polarization for which we can only assign  a dubious membership. Because this tool has been used in previous work, our opinion is that polarimetry can be as useful as photometry to identifications or, in most of the cases, a very useful complement.

Members identified by applying this method are shown in both Figs. 4 and 5 using filled (for member) and open (for non-member) symbols. Circles are used for stars, squares for supergiants, and  starred points  for blue stragglers. Small symbols of any kind denote stars with intrinsic polarization.
We could add an important number of new members to the previous list, and we have also been able to review the situation for other stars listed in the photometric investigations. The last column of Table 2 lists our conclusions.

In particular, five of the observed blue straggler stars are behind the dust located in front of the member stars, but \#195 may be a possible member because of its position in the $ Q_{v}\  versus \  U_{v}$ plot.
As mentioned in the previous section, intrinsic polarization was detected for the star that could explain the position in Fig. 4.
We also confirmed that the red giants \# 116 and 227 are members of the open cluster, as asserted in the literature.
To compare our proposed polarimetric memberships with those coming from other methods, we used the works of Mermilliod et al. (\cite{MMU08}) and  Frinchaboy \& Majewski (\cite{FM08}). In the first investigation, radial velocities of giant stars in the region of a cluster are compared with the cluster mean velocity to assign membership. We have four red giants in common with that study (\#55, 116, 227, 347) and our membership results are in agreement. In the second investigation, the star proper motion (from Hipparcos and Tycho-2 catalogues), radial velocity, and spatial distribution are combined to detect cluster members.
Among a group of seven stars in common (\#55, 116, 180, 202, 227, 342, 347), there are membership discrepancies for only star \# 202 ($ P_{max}$=4.83$\%,  \theta_{v}$= 76$\fdg1$), for which they find a probability of 51.2$\%$ on the basis of its radial velocity but a 0$\%$ probability using the proper motions. According to both Figs. 4 and 5, the star appears to be located behind the same sheet of dust as the remaining members, and for that we consider the star as a member, as the photometric plots suggest. Column 6 in Table 2 lists the memberships determined in these two works.

In Fig. 4, stars \#187, 270, and 333 are located far away from the member group. The first star
is considered to be part of the cluster by some photometric studies but no membership information is provided by Mermilliod et al. (\cite{MMU08}), Frinchaboy \& Majewski (\cite{FM08}), or Dias et al. (\cite{DAFAL06}). The $ \sigma_{1}$ is not indicative of abnormal polarization as to justify its position in the plot; and it can be seen in Fig. 5 that it has a low polarization relative to the remaining members.  We propose that \#187 is a frontside star, observed projected onto the central core of NGC 5617.
The other two stars (\#270 and 333) both display intrinsic polarization as mentioned in the following section.

To derive mean values of polarization and polarization angle, we used 16 stars with similar Stokes parameters and free of intrinsic polarization: \#79, 80, 93, 94, 100, 106, 116, 169, 180, 185, 186, 214, 225, 276, 294, and 302. We obtained $ P_{max}$ = 4.40$\%$ and $ \theta_{v}$ = 73$\fdg$1$\ \pm$0.9 (both of them being the mean values for all 16).
The mean $ \lambda_{max}$ amounts to 0.53 $\pm$ 0.03 $\mu$m, the value associated with the ISM.

\subsection{Polarization efficiency}

It is known that for the interstellar medium the polarization efficiency (ratio of the maximum amount of polarization to visual extinction) rarely exceeds the empirical upper limit,
     $$P_{\rm max} \leq~3~A_{v}~\simeq~3~R_{v}~E_{B-V}~~~~~   (2)$$
obtained for interstellar dust particles (Hiltner 1956). The polarization efficiency indicates how much polarization is obtained for a
certain amount of extinction and depends mainly on both the alignment efficiency and the magnetic field strength, and also on the amount of depolarization due
to radiation traversing more than one cloud with different field directions.

Figure 5 shows the plot of $ P_{max}\ vs.\ E_{B-V}$. The individual excesses $ E_{B-V}$ were obtained either from the literature or by dereddening the colors and using the relationship between either spectral type and color indexes (Schmidt-Kaler \cite{SK82}). For stars with only photographic UBV measures, the calculated excesses may well be in error.

It can be seen that, apart from five stars (from top to bottom: \#333, 294, 261, 136, and 269), the remainder are located to the right of the interstellar maximum line, indicating that their polarizations are mostly due to the ISM. Star \#333 has intrinsic polarization in Table 2 and, based on its position in the plot, we understand it to be a background star. Star \#294, the second from top, has no evident indications of abnormal polarization in its light and we could not find any suitable explanation of the position in this figure. Most probably, the excess (calculated from photographic photometry) is affected by error. Stars \# 261 (a blue struggler) and \#136 are affected by intrinsic polarization as shown in Table 2; with respect to star \#269, the calculated excess may well also be affected by error.
As in Fig. 4, the two stars \#187 and 333 are shown here as non-members. But regarding \#270, we accept that this star is a member, beacuse even when its polarization includes an intrinsic part,  its position in Fig. 5 matches those of the member stars, and its $ P_{max}$ and $ \theta_{v}$ values (4.31 and 73$\fdg$4, respectively) are coincident with the mean values for members.
To calculate the polarization efficiency, we selected a group of 15 stars with $ P_{v}$ in the range from 4.20$ \%$ to 4.95$ \%$, and obtained a polarization efficiency of about 2.44, which is lower than the standard value for the interstellar dust (of about 5). This value indicates a very high efficiency of the dust that polarizes the light from the cluster stars.

  Neckel \& Klare (1980)  computed the interstellar extinction values and distances of more than 11000 O to F stars in the Milky Way. Their figure \# 174 (314$^\circ$, 0$ ^\circ$) shows the variation in Av with increasing distance in the area of NGC 5617,
starting at  1 kpc and showing that the absorption takes values of between 1.2 and 2.4 mag at the position of  the cluster, in good agreement with the Av calculated from (2) for a $ P_{max}$ of 4.4$ \%$.  The nearest of our stars to the Sun is \#187 (non-member) with a $ P_{v}$= 1.83$ \%$ and a distance of about 1.2 kpc from us, which it implies that not in the Local but in the Carina-Sagittarius arm, in addition to the remaining non-members and the cluster itself.
As can be seen in Fig. 5, there is a scatter in polarization values for the members of NGC 5617, which could be caused by either intracluster dust or the uneven distribution of dust in front of the cluster's face, as seen in any plate of the object. Since NGC 5617 is an intermediate-age open cluster ($8.2  10^7$ yr), we favor the second explanation of the scatter.

\section{Summary}

We have measured the linear multicolor polarization of 72 stars in the region of the open cluster NGC 5617. By analyzing all of these data, we have found that between the Sun's position and the Carina-Sagittarius arm, there is a large region of transparency and that NGC 5617 is located deep inside the arm, or even at its farthest border. In addition, we have found the polarimetric efficiency of the dust in front of the cluster to be very high relative to  the mean value attributed to the ISM.

Polarimetry has proven to be an excellent tool in the task of membership identification. It has been employed in previous investigations where the main goal was the determination of general characteristics of the cluster rather than the precise assessment of membership. The identification studies of members and non-members now typically covers a wider area of the cluster and reaches fainter magnitudes. A comparison of  the polarimetric  and kinematical  memberships of stars in common with other investigations, has confirmed that the
polarimetric observations could help resolve these issues.

\section*{Acknowledgments}

This research has made use of the WEBDA database, operated at the Institute for Astronomy of the University of Vienna. 
We wish to acknowledge the technical support and hospitality at CASLEO during the observing runs.  We also acknowledge the use of the Torino Photopolarimeter
built at Osservatorio Astronomico di Torino (Italy) and operated under
agreement between Complejo Astron\'omico El Leoncito and Osservatorio Astronomico di Torino.
We thanks the anonymous referee for the help to improve the paper. Special thanks go to Dr. Hugo G. Marraco for his useful comments and also to Mrs. M. C. Fanjul de Correbo
for the technical
assistance.




\begin{table}
\caption {Polarimetric Observations of stars in NGC 5617}

\begin{tabular}{lccccc}
\noalign{\smallskip}
\hline \hline
\noalign{\smallskip}

 & U &   B & V &   R &  I \\

Star Id. & $P_{U} (\%) \pm \epsilon_{P_U} $  &  $P_{B} (\%)\pm \epsilon_{P_B}$  &    $P_{V}(\%)\pm  \epsilon_{P_V}$  &  $P_{R}(\%) \pm  \epsilon_{P_R}$  &   $P_{I}(\%  \pm \epsilon_{P_I}$   \\
        & $\theta_{U} (^\circ)\pm  \epsilon_{\theta_U} $ &$\theta_{B} (^\circ)\pm  \epsilon_{\theta_B}$ &$\theta_{V} (^\circ) \pm \epsilon_{\theta_V}$ &$\theta_{R}(^\circ) \pm  \epsilon_{\theta_R}$ &$\theta_{I} (^\circ)\pm  \epsilon_{\theta_I}$ \\


\noalign{\smallskip}
\hline
\noalign{\smallskip}

36& 
  -          -  &
3.15 $\pm$ 0.39 &  
3.00 $\pm$ 0.32 &  
3.33 $\pm$ 0.22 & 
3.11 $\pm$ 0.41 \\
  &  
  -          -   &
70.2 $\pm$  3.5  &   
69.5 $\pm$  3.0  &   
63.1 $\pm$  1.9  &   
61.5 $\pm$  3.7  \\
\noalign{\smallskip}

55&
  -          -  &
2.76 $\pm$ 0.39 &  
2.09 $\pm$ 0.27 &  
2.14 $\pm$ 0.28 &  
1.90 $\pm$ 0.25\\
  &
  -          - &
64.3 $\pm$  4.0&  
60.5 $\pm$  3.7&   
63.3 $\pm$  3.7&   
60.7 $\pm$  3.7 \\ 
\noalign{\smallskip}

58& 
3.40 $\pm$ 0.47 &  
3.51 $\pm$ 0.46 &  
2.11 $\pm$ 0.33 & 
2.54 $\pm$ 0.36 &  
1.67 $\pm$ 0.45 \\
  & 
76.3 $\pm$  3.9  & 
82.2 $\pm$  3.7  &   
94.3 $\pm$  4.4  &  
92.0 $\pm$  4.0  &   
93.5 $\pm$  7.5  \\  
\noalign{\smallskip}  

65& 
  -          -   &
  -          -   &
1.79 $\pm$ 0.20  &  
1.69 $\pm$ 0.20  &  
2.08 $\pm$ 0.30 \\  
  & 
  -          -   &
  -          -   &
70.5 $\pm$  3.2  & 
76.9 $\pm$  3.4  &   
72.6 $\pm$  4.1  \\
\noalign{\smallskip}

68*& 
3.01 $\pm$ 0.42 & 
4.27 $\pm$ 0.35 &  
4.21 $\pm$ 0.21 &  
4.19 $\pm$ 0.11 &  
3.87 $\pm$ 0.25 \\
   & 
72.2 $\pm$  4.0  &
73.4 $\pm$  2.3  &   
69.8 $\pm$  1.4  &  
72.3 $\pm$  0.8  &  
71.6 $\pm$  1.8  \\
\noalign{\smallskip}
 
69& 
 -          -   &
 -          -   &
3.19 $\pm$ 0.43 &  
4.01 $\pm$ 0.36 &  
2.51 $\pm$ 0.32 \\
  & 
  -          -   &
  -          -   &
66.8 $\pm$  3.8  & 
70.3 $\pm$  2.6  &  
68.3 $\pm$  3.6  \\
\noalign{\smallskip} 

79& 
 -          -   &
3.41 $\pm$ 0.39 &  
4.81 $\pm$ 0.17 &  
4.95 $\pm$ 0.24 &  
4.24 $\pm$ 0.49 \\
  &
 -           -  &
73.5 $\pm$  3.3 & 
72.8 $\pm$  1.0 &   
71.2 $\pm$  1.4 & 
71.5 $\pm$  3.3 \\
\noalign{\smallskip}

80& 
 -          -   & 
 -          -   &  
4.63 $\pm$ 0.36 &  
4.42 $\pm$ 0.22 &  
4.13 $\pm$ 0.38 \\
  &
  -          -   &
  -          -   &
74.0 $\pm$  2.2  & 
73.8 $\pm$  1.4  &   
66.8 $\pm$  2.6   \\  
\noalign{\smallskip}   
   
93& 
3.95 $\pm$ 0.55 &  
4.86 $\pm$ 0.65 &  
4.55 $\pm$ 0.37 &  
4.10 $\pm$ 0.31 &  
4.79 $\pm$ 0.64 \\
  & 
77.7 $\pm$  4.0  &
78.9 $\pm$  3.8  &  
73.4 $\pm$  2.3  &   
72.5 $\pm$  2.2  &  
77.0 $\pm$  3.8  \\
\noalign{\smallskip}
 
94& 
3.89 $\pm$ 0.44 &  
4.66 $\pm$ 0.45 &   
4.22 $\pm$ 0.26 &    
4.02 $\pm$ 0.18 &  
3.35 $\pm$ 0.43 \\ 
  &
57.1 $\pm$  3.2 & 
72.8 $\pm$  2.8 &  
74.3 $\pm$  1.8 &  
74.2 $\pm$  1.3 &  
70.4 $\pm$  3.7 \\
\noalign{\smallskip}

100&
5.47 $\pm$ 0.47 &  
5.02 $\pm$ 0.37 &  
4.31 $\pm$ 0.29 &  
4.01 $\pm$ 0.29 & 
3.13 $\pm$ 0.46 \\
   & 
64.4 $\pm$  2.5 &
70.7 $\pm$  2.1 &   
73.0 $\pm$  1.9 &   
70.5 $\pm$  2.1 &   
73.1 $\pm$  4.2 \\
\noalign{\smallskip}

106&
  -          -  & 
5.01 $\pm$ 0.67 &  
4.51 $\pm$ 0.20 &     
4.03 $\pm$ 0.21 &     
3.81 $\pm$ 0.45 \\
   &
  -          -   &
81.1 $\pm$  3.8  &  
74.5 $\pm$  1.3  &   
80.1 $\pm$  1.5  &   
78.8 $\pm$  3.4  \\   
\noalign{\smallskip}

107&
  -          -  &  
4.35 $\pm$ 0.25 &
4.18 $\pm$ 0.19 & 
4.44 $\pm$ 0.09 &  
3.60 $\pm$ 0.17 \\
   & 
  -          -  &
76.3 $\pm$  1.6 & 
75.9 $\pm$  1.3 &  
76.5 $\pm$  0.6 &  
76.0 $\pm$  1.3 \\
\noalign{\smallskip}

116*& 
3.72 $\pm$ 0.45 &  
3.71 $\pm$ 0.11 &  
4.15 $\pm$ 0.03 & 
3.94 $\pm$ 0.03 &  
3.55 $\pm$ 0.02 \\
  & 
74.2 $\pm$  3.4 & 
72.5 $\pm$  0.8 &   
71.4 $\pm$  0.2 &   
71.4 $\pm$  0.2 &   
70.7 $\pm$  0.2 \\
\noalign{\smallskip}

136& 
  -           - & 
4.38 $\pm$ 0.48 &  
3.59 $\pm$ 0.19 &  
4.11 $\pm$ 0.22 & 
3.44 $\pm$ 0.34 \\
    &
  -          -   &
72.4 $\pm$  3.1  &  
79.2 $\pm$  1.5  &   
76.4 $\pm$  1.5  &   
80.3 $\pm$  2.8  \\
\noalign{\smallskip}

137&
1.75 $\pm$ 0.28 & 
0.65 $\pm$ 0.25 &  
3.00 $\pm$ 0.21 & 
2.88 $\pm$ 0.24 &  
2.15 $\pm$ 0.30 \\
   &
70.1 $\pm$  4.5  & 
76.2 $\pm$ 10.5  &   
70.7 $\pm$  2.0  &  
72.7 $\pm$  2.4  &   
75.8 $\pm$  4.0  \\
\noalign{\smallskip}     

139& 
4.46 $\pm$ 0.46 & 
4.17 $\pm$ 0.37 & 
4.06 $\pm$ 0.24 & 
4.18 $\pm$ 0.27 &  
3.74 $\pm$ 0.46 \\
   & 
76.8 $\pm$  2.9 &
76.3 $\pm$  2.5 &  
74.9 $\pm$  1.7 &   
76.9 $\pm$  1.8 &  
69.8 $\pm$  3.5 \\
\noalign{\smallskip}

\hline
\end{tabular}
\end{table}

\addtocounter{table}{-1}%
\begin{table}
\caption {Polarimetric Observations of stars in NGC 5617}

\begin{tabular}{lccccc}
\noalign{\smallskip}
\hline \hline
\noalign{\smallskip}

 & U &   B & V &   R &  I \\

Star Id. & $P_{U} (\%) \pm \epsilon_{P_U} $  &  $P_{B} (\%)\pm \epsilon_{P_B}$  &    $P_{V}(\%)\pm  \epsilon_{P_V}$  &  $P_{R}(\%) \pm  \epsilon_{P_R}$  &   $P_{I}(\%  \pm \epsilon_{P_I}$   \\
        & $\theta_{U} (^\circ)\pm  \epsilon_{\theta_U} $ &$\theta_{B} (^\circ)\pm  \epsilon_{\theta_B}$ &$\theta_{V} (^\circ) \pm \epsilon_{\theta_V}$ &$\theta_{R}(^\circ) \pm  \epsilon_{\theta_R}$ &$\theta_{I} (^\circ)\pm  \epsilon_{\theta_I}$ \\


\noalign{\smallskip}
\hline
\noalign{\smallskip}

143& 
  -          -  &
3.89 $\pm$ 0.39 &
2.84 $\pm$ 0.30 &  
3.24 $\pm$ 0.31 & 
3.13 $\pm$ 0.42 \\
   & 
  -          -  &
71.9 $\pm$  2.9 &  
72.3 $\pm$  3.0 &  
72.5 $\pm$  2.7 &  
66.7 $\pm$  3.8 \\
\noalign{\smallskip}

146& 
3.90 $\pm$ 0.36 &  
4.33 $\pm$ 0.38 &    
3.88 $\pm$ 0.24 &     
3.90 $\pm$ 0.33 &  
4.71 $\pm$ 0.40 \\
   & 
74.6 $\pm$  2.6 & 
68.5 $\pm$  2.5 &  
69.2 $\pm$  1.8 &  
67.9 $\pm$  2.4 & 
70.3 $\pm$  2.4 \\
\noalign{\smallskip}

149& 
3.32 $\pm$ 0.40 & 
1.19 $\pm$ 0.25 &  
1.49 $\pm$ 0.19 & 
1.45 $\pm$ 0.19 &  
1.40 $\pm$ 0.21 \\  
   & 
89.8 $\pm$  3.4  & 
91.8 $\pm$  5.9  &  
74.2 $\pm$  3.6  &   
75.6 $\pm$  3.7  &   
66.6 $\pm$  4.3  \\
\noalign{\smallskip}

154& 
3.46 $\pm$ 0.41 & 
1.93 $\pm$ 0.14 &  
3.14 $\pm$ 0.18 &  
2.61 $\pm$ 0.18 &  
2.03 $\pm$ 0.16 \\
  & 
68.8 $\pm$  3.4  & 
69.1 $\pm$  2.1  &   
66.5 $\pm$  1.6  &   
64.4 $\pm$  2.0  &  
69.5 $\pm$  2.2 \\
\noalign{\smallskip}

169& 
4.60 $\pm$ 0.51 & 
4.64 $\pm$ 0.36 &  
4.40 $\pm$ 0.19 & 
4.37 $\pm$ 0.22 &  
3.50 $\pm$ 0.42 \\
   & 
70.9 $\pm$  3.2  &
65.8 $\pm$  2.2  &   
73.7 $\pm$  1.2  &   
72.5 $\pm$  1.4  &   
74.9 $\pm$  3.4  \\
\noalign{\smallskip}

175&
3.66 $\pm$ 0.43 & 
4.24 $\pm$ 0.35 &  
4.14 $\pm$ 0.24 & 
4.22 $\pm$ 0.17 &  
3.85 $\pm$ 0.41 \\
   &
73.0 $\pm$  3.3 & 
81.1 $\pm$  2.4 &  
80.1 $\pm$  1.7 &  
80.1 $\pm$  1.2 &   
80.0 $\pm$  3.0 \\
\noalign{\smallskip}

176& 
  -          -  &  
2.71 $\pm$ 0.32 &  
2.50 $\pm$ 0.21 &  
2.33 $\pm$ 0.13 &    
2.13 $\pm$ 0.19 \\
   & 
  -          -   &  
65.5 $\pm$  3.4  &  
65.1 $\pm$  2.4  &   
65.1 $\pm$  1.6  &   
68.2 $\pm$  2.5  \\
\noalign{\smallskip}

180& 
4.25 $\pm$ 0.17 &  
4.48 $\pm$ 0.24 &  
4.66 $\pm$ 0.35 &  
4.49 $\pm$ 0.17 &  
3.97 $\pm$ 0.25 \\
   & 
70.7 $\pm$  1.1  &
74.4 $\pm$  1.5  &  
73.5 $\pm$  2.1  &  
73.1 $\pm$  1.1  &  
73.5 $\pm$  1.8 \\
\noalign{\smallskip}

182&
4.48 $\pm$ 0.55 & 
3.98 $\pm$ 0.39 &  
3.91 $\pm$ 0.26 & 
3.73 $\pm$ 0.18 &  
2.94 $\pm$ 0.37 \\
 &
74.1 $\pm$  3.5  & 
75.6 $\pm$  2.8  &   
74.7 $\pm$  1.9  &   
74.9 $\pm$  1.4  &  
73.8 $\pm$  3.6  \\
\noalign{\smallskip}

184&
4.89 $\pm$ 0.62 & 
4.66 $\pm$ 0.37 & 
4.56 $\pm$ 0.21 & 
4.52 $\pm$ 0.25 & 
4.20 $\pm$ 0.37 \\
   &  
75.4 $\pm$  3.6  & 
76.1 $\pm$  2.3  &   
75.4 $\pm$  1.3  &   
76.2 $\pm$  1.6  &   
69.6 $\pm$  2.5  \\
\noalign{\smallskip}

185*& 
3.95 $\pm$ 0.13 &  
4.21 $\pm$ 0.23 &  
4.44 $\pm$ 0.13 & 
4.16 $\pm$ 0.21 &  
3.63 $\pm$ 0.20 \\
& 
72.2 $\pm$  0.9 &
74.8 $\pm$  1.6 &   
74.8 $\pm$  0.8 &   
74.3 $\pm$  1.4 &   
72.8 $\pm$  1.6 \\
\noalign{\smallskip}

186& 
4.30 $\pm$ 0.55 &  
4.38 $\pm$ 0.42 &  
4.67 $\pm$ 0.22 & 
4.59 $\pm$ 0.24 &  
4.22 $\pm$ 0.52 \\
   & 
72.7 $\pm$  3.6  &
76.6 $\pm$  2.7  &   
72.3 $\pm$  1.3  &  
74.5 $\pm$  1.5  &   
72.9 $\pm$  3.5  \\
\noalign{\smallskip}

187* & 
1.64 $\pm$ 0.54 & 
2.18 $\pm$ 0.38 & 
1.37 $\pm$ 0.23 &  
2.11 $\pm$ 0.25 &  
1.71 $\pm$ 0.52 \\
     &
76.3 $\pm$  9.1&
68.5 $\pm$  4.9&  
70.4 $\pm$  4.8&  
55.2 $\pm$  3.4&   
76.2 $\pm$  8.4  \\
\noalign{\smallskip}

188&
  -          -  & 
3.58 $\pm$ 0.53 & 
4.95 $\pm$ 0.45 &  
4.46 $\pm$ 0.49 & 
4.18 $\pm$ 0.60 \\
   &
  -          -   &
71.2 $\pm$  4.2  & 
74.4 $\pm$  2.6  &   
76.1 $\pm$  3.1  &   
72.2 $\pm$  4.1  \\
\noalign{\smallskip}

190& 
3.16 $\pm$ 0.47 &  
4.20 $\pm$ 0.33 &    
3.98 $\pm$ 0.18 &    
3.93 $\pm$ 0.10 &    
3.20 $\pm$ 0.46 \\
   &
75.4 $\pm$  4.2 & 
73.6 $\pm$  2.2 &  
73.4 $\pm$  1.3 &   
74.1 $\pm$  0.7 &  
72.0 $\pm$  4.1 \\
\noalign{\smallskip}

195* &  
5.28 $\pm$ 0.47 & 
4.13 $\pm$ 0.43 &  
3.06 $\pm$ 0.46 & 
4.42 $\pm$ 0.48 &  
3.18 $\pm$ 0.44 \\
&
79.9 $\pm$  2.5& 
71.1 $\pm$  3.0& 
65.1 $\pm$  4.3&  
66.8 $\pm$  3.1&
60.2 $\pm$  3.9   \\
\noalign{\smallskip}

196& 
3.90 $\pm$ 0.78 & 
2.80 $\pm$ 0.24 & 
2.01 $\pm$ 0.51 & 
2.35 $\pm$ 0.42 &  
2.69 $\pm$ 0.61 \\
   & 
54.4 $\pm$  5.6  &
73.9 $\pm$  2.4  &   
78.9 $\pm$  7.1  &   
70.7 $\pm$  5.1  &  
68.0 $\pm$  6.4 \\
\noalign{\smallskip}

202*&
4.72 $\pm$ 0.31 &
4.84 $\pm$ 0.15 & 
4.63 $\pm$ 0.11 & 
4.50 $\pm$ 0.10 &  
3.89 $\pm$ 0.23 \\
 & 
71.3 $\pm$  1.9 &  
75.4 $\pm$  0.9 &  
76.0 $\pm$  0.7 &   
76.1 $\pm$  0.6 &   
73.9 $\pm$  1.7 \\
\noalign{\smallskip}

203&
4.20 $\pm$ 0.35 & 
4.33 $\pm$ 0.23 &  
4.03 $\pm$ 0.58 &  
4.33 $\pm$ 0.47 & 
3.81 $\pm$ 0.36 \\
   & 
70.0 $\pm$  2.4 &
69.3 $\pm$  1.5 &
74.1 $\pm$  4.1 &
73.7 $\pm$  3.1 &
70.8 $\pm$  2.7 \\
\noalign{\smallskip}

205* & 
5.41 $\pm$ 1.24 &  
4.71 $\pm$ 0.43 & 
4.00 $\pm$ 0.34 &  
4.00 $\pm$ 0.36 &
3.14 $\pm$ 0.50  \\
     & 
68.0 $\pm$  6.4&
71.5 $\pm$  2.6&  
75.8 $\pm$  2.4& 
75.6 $\pm$  2.6& 
84.4 $\pm$  4.5 \\
\noalign{\smallskip}

\hline
\end{tabular}
\end{table}

\addtocounter{table}{-1}%
\begin{table}
\caption {Polarimetric Observations of stars in NGC 5617}

\begin{tabular}{lccccc}
\noalign{\smallskip}
\hline \hline
\noalign{\smallskip}

 & U &   B & V &   R &  I \\

Star Id. & $P_{U} (\%) \pm \epsilon_{P_U} $  &  $P_{B} (\%)\pm \epsilon_{P_B}$  &    $P_{V}(\%)\pm  \epsilon_{P_V}$  &  $P_{R}(\%) \pm  \epsilon_{P_R}$  &   $P_{I}(\%  \pm \epsilon_{P_I}$   \\
        & $\theta_{U} (^\circ)\pm  \epsilon_{\theta_U} $ &$\theta_{B} (^\circ)\pm  \epsilon_{\theta_B}$ &$\theta_{V} (^\circ) \pm \epsilon_{\theta_V}$ &$\theta_{R}(^\circ) \pm  \epsilon_{\theta_R}$ &$\theta_{I} (^\circ)\pm  \epsilon_{\theta_I}$ \\


\noalign{\smallskip}
\hline
\noalign{\smallskip}

207* & 
3.41 $\pm$ 0.52 &  
4.06 $\pm$ 0.38 &  
3.48 $\pm$ 0.21 &  
3.35 $\pm$ 0.39 &  
3.35 $\pm$ 0.44 \\
     & 
64.3 $\pm$   4.3& 
66.7 $\pm$   2.7&  
78.9 $\pm$   1.7&  
74.9 $\pm$   3.3& 
72.9 $\pm$   3.7   \\
\noalign{\smallskip}

214&
3.16 $\pm$ 0.38 &
4.16 $\pm$ 0.32 &
4.15 $\pm$ 0.16 &
4.17 $\pm$ 0.18 &
3.82 $\pm$ 0.26 \\
   & 
72.7 $\pm$  3.4  & 
73.5 $\pm$  2.2  &   
71.0 $\pm$  1.1  &   
73.2 $\pm$  1.2  &  
69.7 $\pm$  1.9  \\
\noalign{\smallskip}

215&
  -          -  & 
3.83 $\pm$ 0.52 &  
3.35 $\pm$ 0.15 &  
3.40 $\pm$ 0.13 &  
2.97 $\pm$ 0.17 \\
   &
  -          -   &
74.4 $\pm$  3.9  &
69.2 $\pm$  1.3  &   
72.2 $\pm$  1.1  &   
70.7 $\pm$  1.6  \\
\noalign{\smallskip}

216* &
4.38 $\pm$ 0.41 &  
4.80 $\pm$ 0.39 &  
4.22 $\pm$ 0.24 & 
4.05 $\pm$ 0.12 &  
3.48 $\pm$ 0.29 \\
     &
71.0 $\pm$  2.7 &
71.1 $\pm$  2.3 & 
74.5 $\pm$  1.6 & 
73.9 $\pm$  0.8 & 
70.6 $\pm$  2.4 \\
\noalign{\smallskip}

221& 
3.60 $\pm$ 0.36 &  
3.84 $\pm$ 0.20 &  
3.49 $\pm$ 0.16 &  
3.34 $\pm$ 0.19 & 
3.02 $\pm$ 0.27 \\
& 
74.9 $\pm$  2.9  &
76.1 $\pm$  1.5  &   
76.5 $\pm$  1.3  &   
74.2 $\pm$  1.6  &   
73.2 $\pm$  2.6  \\
\noalign{\smallskip}

225*& 
4.83 $\pm$ 0.56 &
4.60 $\pm$ 0.37 & 
4.47 $\pm$ 0.32 &  
4.56 $\pm$ 0.19 &  
3.72 $\pm$ 0.44 \\
    & 
72.3 $\pm$  3.3 & 
71.9 $\pm$  2.3 &   
73.6 $\pm$  2.0 &  
76.2 $\pm$  1.2 &  
75.1 $\pm$  3.4 \\
\noalign{\smallskip}

226*& 
  -          -  &
  -          -  &
4.00 $\pm$ 0.29 &  
3.72 $\pm$ 0.34 & 
4.78 $\pm$ 0.54 \\
    &
  -          -  &
  -          -  & 
76.9 $\pm$  2.1 & 
77.0 $\pm$  2.6 &  
70.9 $\pm$  3.2 \\
\noalign{\smallskip}

227*& 
7.90 $\pm$ 1.00 & 
4.36 $\pm$ 0.15 &    
4.24 $\pm$ 0.14 &   
3.87 $\pm$ 0.16 &   
3.43 $\pm$ 0.14 \\
  & 
73.4 $\pm$  3.6 &
78.7 $\pm$  1.0 &  
73.2 $\pm$  0.9 & 
74.3 $\pm$  1.2 &  
71.7 $\pm$  1.2 \\
\noalign{\smallskip}

238& 
2.75 $\pm$ 0.56 &  
4.03 $\pm$ 0.42 &  
4.13 $\pm$ 0.44 &  
4.57 $\pm$ 0.42 &  
2.16 $\pm$ 0.65 \\
&
65.1 $\pm$  5.7  &
78.6 $\pm$  3.0  &   
78.3 $\pm$  3.0  &   
69.7 $\pm$  2.6  &   
74.7 $\pm$  8.4  \\
\noalign{\smallskip}

244& 
3.52 $\pm$ 0.43 &
3.53 $\pm$ 0.45 &  
3.78 $\pm$ 0.25 &  
3.48 $\pm$ 0.40 &  
3.61 $\pm$ 0.41 \\
   & 
85.4 $\pm$  3.5 & 
79.1 $\pm$  3.6 &  
74.8 $\pm$  1.9 &  
81.8 $\pm$  3.3 &  
74.1 $\pm$  3.2 \\
\noalign{\smallskip}

255& 
  -          -  &
3.52 $\pm$ 0.37 & 
3.33 $\pm$ 0.41 &  
2.45 $\pm$ 0.19 &  
3.66 $\pm$ 0.40 \\
   &
  -          -  & 
69.8 $\pm$  3.0 &
78.6 $\pm$  3.5 &   
73.5 $\pm$  2.2 &
72.4 $\pm$  3.1 \\
\noalign{\smallskip}

258*&
3.78 $\pm$ 0.38 &  
4.19 $\pm$ 0.30 &  
3.62 $\pm$ 0.17 &  
3.91 $\pm$ 0.10 &  
3.18 $\pm$ 0.35 \\
    &
77.6 $\pm$  2.9  & 
76.3 $\pm$  2.0  &   
76.8 $\pm$  1.3  &   
77.2 $\pm$  0.7  &   
78.2 $\pm$  3.1  \\
\noalign{\smallskip}

259& 
  -          -  & 
4.71 $\pm$ 0.59 & 
3.62 $\pm$ 0.39 &  
3.28 $\pm$ 0.32 &  
3.00 $\pm$ 0.37  \\
   & 
  -          -   &  
71.8 $\pm$  3.6  & 
68.4 $\pm$  3.1  &   
69.0 $\pm$  2.8  &   
55.5 $\pm$  3.5  \\
\noalign{\smallskip}

260& 
  -          -  &
4.27 $\pm$ 0.45 &  
2.91 $\pm$ 0.19 &  
3.15 $\pm$ 0.27 &  
2.98 $\pm$ 0.33  \\
   & 
  -          -   &
60.7 $\pm$  3.0  & 
67.0 $\pm$  1.9  &   
68.4 $\pm$  2.4  &   
71.9 $\pm$  3.2  \\
\noalign{\smallskip}   

261&
3.97 $\pm$ 0.24 & 
3.86 $\pm$ 0.11 &  
3.70 $\pm$ 0.11 &  
3.69 $\pm$ 0.12 & 
3.62 $\pm$ 0.12 \\
  & 
70.2 $\pm$  1.7 & 
76.5 $\pm$  0.8 &  
74.1 $\pm$  0.9 &  
72.6 $\pm$  0.9 &  
66.9 $\pm$  0.9 \\
\noalign{\smallskip}

269& 
  -         -   &
  -         -   &
3.36 $\pm$ 0.26 &  
3.43 $\pm$ 0.16 &  
2.99 $\pm$ 0.27  \\
   & 
  -          -   & 
  -          -   &
68.8 $\pm$  2.2  & 
68.7 $\pm$  1.3  &   
69.8 $\pm$  2.6  \\
 \noalign{\smallskip}  

270& 
5.06 $\pm$ 0.34 &  
4.24 $\pm$ 0.24 &  
3.94 $\pm$ 0.14 &  
3.97 $\pm$ 0.16 & 
3.39 $\pm$ 0.14 \\
   & 
73.2 $\pm$  1.9  & 
74.8 $\pm$  1.6  &   
73.4 $\pm$  1.0  &  
73.3 $\pm$  1.2  &  
70.6 $\pm$  1.2  \\
\noalign{\smallskip}

274& 
5.27 $\pm$ 0.62 &  
3.21 $\pm$ 0.43 &  
3.76 $\pm$ 0.27 &  
4.57 $\pm$ 0.27 &  
3.55 $\pm$ 0.49 \\
   & 
69.3 $\pm$  3.3  &
72.6 $\pm$  3.8  &   
75.7 $\pm$  2.1  &   
73.6 $\pm$  1.7  &  
66.3 $\pm$  3.9 \\
\noalign{\smallskip}

275& 
  -          -  & 
4.31 $\pm$ 0.51 & 
3.51 $\pm$ 0.45 &  
4.35 $\pm$ 0.46 &  
2.67 $\pm$ 0.52 \\
   &
  -          -   &
65.4 $\pm$  3.4  &  
73.3 $\pm$  3.6  & 
63.9 $\pm$  3.0  &  
66.6 $\pm$  5.5  \\
\noalign{\smallskip}

\noalign{\smallskip}
\hline
\end{tabular}
\end{table}

\addtocounter{table}{-1}%
\begin{table}
\caption {Polarimetric Observations of stars in NGC 5617}

\begin{tabular}{lccccc}
\noalign{\smallskip}
\hline \hline
\noalign{\smallskip}

 & U &   B & V &   R &  I \\

Star Id. & $P_{U} (\%) \pm \epsilon_{P_U} $  &  $P_{B} (\%)\pm \epsilon_{P_B}$  &    $P_{V}(\%)\pm  \epsilon_{P_V}$  &  $P_{R}(\%) \pm  \epsilon_{P_R}$  &   $P_{I}(\%  \pm \epsilon_{P_I}$   \\
        & $\theta_{U} (^\circ)\pm  \epsilon_{\theta_U} $ &$\theta_{B} (^\circ)\pm  \epsilon_{\theta_B}$ &$\theta_{V} (^\circ) \pm \epsilon_{\theta_V}$ &$\theta_{R}(^\circ) \pm  \epsilon_{\theta_R}$ &$\theta_{I} (^\circ)\pm  \epsilon_{\theta_I}$ \\


\noalign{\smallskip}
\hline
\noalign{\smallskip}

276& 
3.81 $\pm$ 0.46 & 
4.29 $\pm$ 0.33 & 
4.52 $\pm$ 0.36 &  
4.66 $\pm$ 0.31 & 
4.33 $\pm$ 0.49 \\
   & 
68.5 $\pm$  3.4  & 
74.8 $\pm$  2.2  &   
72.7 $\pm$  2.3  &  
73.4 $\pm$  1.9  &   
76.9 $\pm$  3.2  \\
\noalign{\smallskip}

292& 
4.76 $\pm$ 0.39 & 
4.22 $\pm$ 0.39 &  
3.75 $\pm$ 0.19 & 
3.67 $\pm$ 0.13 & 
2.61 $\pm$ 0.31 \\
   & 
71.1 $\pm$  2.3  & 
72.9 $\pm$  2.6  &   
78.6 $\pm$  1.4  &   
74.5 $\pm$  1.0  &   
75.2 $\pm$  3.4  \\
\noalign{\smallskip}

294&
3.63 $\pm$ 0.52 &  
3.58 $\pm$ 0.37 &  
4.14 $\pm$ 0.36 &
3.56 $\pm$ 0.19 &  
2.68 $\pm$ 0.42 \\
   & 
58.6 $\pm$  4.1  & 
65.2 $\pm$  2.9  &   
72.0 $\pm$  2.5  &   
72.4 $\pm$  1.5  &   
76.1 $\pm$  4.4  \\
\noalign{\smallskip}

295&
  -          -  & 
  -          -  & 
3.79 $\pm$ 0.24 & 
3.61 $\pm$ 0.22 &  
3.53 $\pm$ 0.40  \\
   &
  -          -   &
  -          -   & 
74.8 $\pm$  1.8  &  
75.7 $\pm$  1.7  &   
75.1 $\pm$  3.2  \\
 \noalign{\smallskip}

302& 
3.03 $\pm$ 0.47 & 
3.96 $\pm$ 0.49 & 
4.15 $\pm$ 0.25 & 
4.48 $\pm$ 0.19 &  
3.76 $\pm$ 0.38  \\
   &
72.9 $\pm$  4.4  & 
69.9 $\pm$  3.5  &   
70.8 $\pm$  1.7  &   
73.1 $\pm$  1.2  &   
74.7 $\pm$  2.9  \\   
\noalign{\smallskip}  

321& 
  -          -  &
  -          -  &  
2.21 $\pm$ 0.15 &  
1.78 $\pm$ 0.19 &  
1.92 $\pm$ 0.68  \\
   & 
  -          -   & 
  -          -   &
65.6 $\pm$  1.9  &  
69.9 $\pm$  3.0  &   
69.7 $\pm$  9.7  \\  
\noalign{\smallskip}   
   
327& 
  -          -  &  
4.06 $\pm$ 0.58 & 
2.88 $\pm$ 0.21 &  
2.89 $\pm$ 0.17 &  
2.71 $\pm$ 0.18  \\
   &
  -          -   &
69.1 $\pm$  4.1  &  
69.7 $\pm$  2.1  &  
68.6 $\pm$  1.7  &   
71.2 $\pm$  1.9  \\   
 \noalign{\smallskip}

330& 
3.93 $\pm$ 0.30 &  
4.48 $\pm$ 0.23 & 
4.04 $\pm$ 0.10 &  
4.22 $\pm$ 0.09 &  
3.81 $\pm$ 0.15  \\
   &
71.5 $\pm$  2.2  & 
70.3 $\pm$  1.5  &   
73.8 $\pm$  0.7  &   
72.6 $\pm$  0.6  &   
73.4 $\pm$  1.1  \\
\noalign{\smallskip}

 331& 
   -          -  &
 2.28 $\pm$ 0.47 &  
 3.46 $\pm$ 0.23 &  
 3.21 $\pm$ 0.18 &  
 3.16 $\pm$ 0.27 \\
    &
   -          -   &
 75.6 $\pm$  5.8  &  
 66.1 $\pm$  1.9  &   
 71.1 $\pm$  1.6  &  
 67.9 $\pm$  2.4  \\ 
 \noalign{\smallskip}  
 
333& 
  -          -  &  
3.91 $\pm$ 0.61 &  
5.79 $\pm$ 0.32 &  
4.84 $\pm$ 0.35 &  
5.13 $\pm$ 0.48 \\
   & 
  -          -   &  
59.4 $\pm$  4.4  &  
71.1 $\pm$  1.6  &   
69.3 $\pm$  2.1  &   
67.2 $\pm$  2.7  \\   
\noalign{\smallskip}    

334& 
4.35 $\pm$ 0.45 &
4.89 $\pm$ 0.32 &  
4.45 $\pm$ 0.18 &  
4.47 $\pm$ 0.18 &  
4.14 $\pm$ 0.38  \\
   &
73.4 $\pm$  2.9  & 
74.9 $\pm$  1.9  &   
77.4 $\pm$  1.2  &   
73.3 $\pm$  1.2  &   
71.5 $\pm$  2.6  \\   
\noalign{\smallskip}   

336&
  -          -  &
3.44 $\pm$ 0.48 &  
3.35 $\pm$ 0.27 &  
3.10 $\pm$ 0.43 &  
4.72 $\pm$ 0.56  \\
   & 
  -          -   &
78.0 $\pm$  4.0  &   
60.1 $\pm$  2.3  &   
66.8 $\pm$  3.9  &   
64.0 $\pm$  3.4  \\
 \noalign{\smallskip}    

339*& 
3.66 $\pm$ 0.46 &  
3.83 $\pm$ 0.41 & 
3.33 $\pm$ 0.38 &    
3.75 $\pm$ 0.27 &  
3.69 $\pm$ 0.42  \\
&
65.6 $\pm$  3.6  & 
70.4 $\pm$  3.1  &   
70.9 $\pm$  3.2  &  
71.8 $\pm$  2.1  &   
66.7 $\pm$  3.2  \\  
\noalign{\smallskip}

342& 
  -          -  &  
3.07 $\pm$ 0.36 &  
2.89 $\pm$ 0.17 &  
3.06 $\pm$ 0.13 & 
2.90 $\pm$ 0.11  \\
   &
  -          -   &
67.7 $\pm$  3.3  &  
65.7 $\pm$  1.7  &  
68.6 $\pm$  1.2  &  
66.7 $\pm$  1.1   \\   
\noalign{\smallskip}   

347&  
  -          -  &
2.17 $\pm$ 0.24 &  
1.89 $\pm$ 0.14 &  
2.25 $\pm$ 0.12 &  
1.85 $\pm$ 0.13 \\
 & 
   -          -   &
 66.0 $\pm$  3.2  & 
 67.1 $\pm$  2.1  &   
 67.3 $\pm$  1.5  &   
 69.1 $\pm$  2.0  \\  
\noalign{\smallskip}

379& 
  -          -  & 
3.44 $\pm$ 0.36 &  
2.97 $\pm$ 0.24 & 
3.18 $\pm$ 0.15 & 
2.96 $\pm$ 0.38 \\
   &
  -          -   &
74.6 $\pm$  3.0  &
74.3 $\pm$  2.3  &   
73.9 $\pm$  1.3  &   
78.8 $\pm$  3.7  \\
\noalign{\smallskip}

\hline
\end{tabular}

\begin{list}{}{}
\item  {$\mathit a: $} ~Identifications from Haug (\cite{H78}); *:Membership according to the literature.
\end{list}
\end{table}  

\newpage
\clearpage
\begin{table}
\caption[]{Polarization results of stars in direction to NGC 5617}
\label{table:2}
\begin{tabular}{lccccccc}
\noalign{\smallskip}
\hline \hline
\noalign{\smallskip}
Star$^{~a}$ & Other ids.$^{~b}$ & $P_{\rm max} \pm \epsilon_{p} $ & $\sigma_1~^{c} $ &
                $\lambda_{\rm max} \pm \epsilon_{\lambda}$ &  kin.~memb.~$^{d}$
		 & this work \\
& & $ \%$&  & $\mu$m &      \\
\noalign{\smallskip}
\hline
\noalign{\smallskip}

36  &L96  &  3.31 $\pm$   0.10&  0.63&  0.63$\pm$   0.05&             &  nm \\ 
55  &L180,RG &  2.36 $\pm$   0.22&  1.04&  0.49$\pm$   0.07& nm(M; FM)    &  nm  \\ 
65  &L98  &  1.87 $\pm$   0.19&  1.25&  0.72$\pm$   0.20&             &  nm \\         
68* &L105,BS &  4.26 $\pm$   0.08&  0.78&  0.61$\pm$   0.03&            & member \\  
79  &L46  &  4.93 $\pm$   0.39&  1.61&  0.70$\pm$   0.12&            &   member \\ 
80  &L158 &  4.61 $\pm$   0.08&  0.29&  0.57$\pm$   0.02&             &     nm \\  
93  &L94  &  4.60 $\pm$   0.22&  1.02&  0.54$\pm$   0.06&            &  member \\  
94  &L45  &  4.37 $\pm$   0.10&  0.59&  0.50$\pm$   0.02&            &   member \\         
100  &L47  &  5.05 $\pm$   0.29&  1.00&  0.43$\pm$   0.03&            &   member \\ 
106  &L91  &  4.56 $\pm$   0.17&  0.71&  0.49$\pm$   0.04&            &  member \\             
107  &L48  &  4.50 $\pm$   0.22&  1.32&  0.55$\pm$   0.06&            &   member \\   
116* &L42  &  4.13 $\pm$   0.03&  1.37&  0.55$\pm$   0.01&  m(M; FM)  &  member \\ 
136  &L89  &  3.91 $\pm$   0.23&  1.80&  0.61$\pm$   0.11&            &    nm    \\          
139  &L39  &  4.36 $\pm$   0.17&  1.10&  0.52$\pm$    0.05&           &  member \\         
143  &L51  &  3.36 $\pm$   0.33&  1.78&  0.55$\pm$   0.13&            &     nm  \\           
146  &L49  &  4.33 $\pm$   0.29&  1.84&  0.58$\pm$   0.08&            &   member \\  
154  &L52  &  2.49 $\pm$   0.27&  3.47&  0.57$\pm$   0.13&            &     nm  \\           
169  &L53  &  4.61 $\pm$   0.15&  0.92&  0.49$\pm$   0.03&            &   member \\  
175  &L15  &  4.31 $\pm$   0.06&  0.48&  0.56$\pm$    0.02&        &  member \\  
176 &L114 &  2.57 $\pm$   0.09&  0.55&  0.51$\pm$   0.03&             &   nm   \\         
180  &L54  &  4.77 $\pm$   0.05&  0.47&  0.53$\pm$   0.01&  m(FM)     &   member \\ 
182  &L86,BS  &  4.18 $\pm$   0.18&  0.87&  0.46$\pm$   0.03&            &  poss.member\\  
184  &L87  &  4.76 $\pm$   0.14&  0.95&  0.53$\pm$   0.04&            &  member \\      
185* &L31,BS  &  4.43 $\pm$   0.03&  0.32&  0.52$\pm$   0.01&           &  member \\
186 &L115 &  4.75 $\pm$   0.07&  0.43&  0.56$\pm$   0.02&             &  member \\     
187* &L2   &  1.83 $\pm$   0.26&  1.52&  0.63$\pm$   0.20&        &  nm   \\
188  &L85  &  4.57 $\pm$   0.30&  1.04&  0.65$\pm$   0.10&            &   member \\     
190  &L16  &  4.07 $\pm$   0.09&  0.83&  0.55$\pm$    0.03&        &  member \\  
195* &L1,BS   &  4.42 $\pm$   0.62&  2.21&  0.44$\pm$   0.09&        &  poss. member\\
196  &L55  &  2.80 $\pm$   0.28&  1.39&  0.45$\pm$   0.10&            &     nm     \\  
202* &L29,BS  &  4.83 $\pm$   0.10&  1.33&  0.50$\pm$   0.02&  nm(FM)   &  member \\  
203  &L56  &  4.53 $\pm$   0.11&  0.65&  0.52$\pm$   0.02&            &   member \\  
205* &L3   &  4.51 $\pm$   0.30&  0.93&  0.44$\pm$   0.05&        &  member \\        
207* &L4   &  3.69 $\pm$   0.16&  1.01&  0.52$\pm$   0.05&        &  member \\ 
214 &L116 &  4.23 $\pm$   0.07&  0.59&  0.60$\pm$   0.02&             &  member \\  
215  &L84  &  3.46 $\pm$   0.11&  0.98&  0.56$\pm$   0.04&            & poss.member  \\          
216* &L5   &  4.33 $\pm$   0.06&  0.51&  0.57$\pm$   0.02&        &  member \\  
221  &L57  &  3.72 $\pm$   0.11&  0.95&  0.49$\pm$   0.03&            &   member \\  
225* &L28  &  4.82 $\pm$   0.17&  0.94&  0.51$\pm$   0.03&           &  member \\         
226* &L26  &  4.16 $\pm$   0.48&  1.91&  0.70$\pm$   0.22&           &  member  \\        
227* &L17,RG  &  4.38 $\pm$   0.18&  2.29&  0.48$\pm$   0.03&  m(M; FM)& member \\         
238  &L59  &  3.95 $\pm$   0.45&  1.48&  0.55$\pm$   0.13&            &   member \\         
244  &L23  &  3.86 $\pm$   0.12&  0.66&  0.55$\pm$    0.04&           &  member \\                     258* &L81  &  4.01 $\pm$   0.17&  1.69&  0.54$\pm$  0.05&            &   member \\ 
259 &L119 &  4.02 $\pm$   0.42&  1.06&  0.45$\pm$   0.06&             &  member \\    
260 &L118 &  3.20 $\pm$   0.30&  2.26&  0.56$\pm$   0.14&             & member  \\             261  &L20,BS  &  3.98 $\pm$   0.14&  2.32&  0.55$\pm$    0.04&        &  poss.member \\

\noalign{\smallskip}
\hline
\end{tabular}
\end{table}  

\addtocounter{table}{-1}%
\begin{table}
\caption{continued}
\begin{tabular}{lccccccc}

\noalign{\smallskip}
\hline\hline
\noalign{\smallskip}

Star$^{~a}$ & Other ids.$^{~b}$& $P_{\rm max} \pm  \epsilon_{p}$ & $\sigma_1~^{c}$ &
        $\lambda_{\rm max} \pm \epsilon_{\lambda}$ &  kin.~memb.~$^{d}$
		 & this work \\

& & $\%$  &  & $\mu$m &   \\

\noalign{\smallskip}
\hline
\noalign{\smallskip}

269 &L120 &  3.46 $\pm$   0.11&  0.59&  0.58$\pm$   0.05&             &   nm \\    
270 & L82  &  4.31 $\pm$   0.26&  2.36&  0.49$\pm$   0.05&            & poss.member  \\      
274 & L62  &  4.19 $\pm$   0.39&  2.39&  0.59$\pm$   0.13&            &   member \\         
275 & L61  &  4.09 $\pm$   0.57&  1.58&  0.47$\pm$   0.11&            &   member \\        
276 & L68  &  4.71 $\pm$   0.05&  0.24&  0.60$\pm$   0.01&            &   member \\        
379 & L64  &  3.25 $\pm$   0.15&  1.11&  0.57$\pm$   0.07&            &     nm   \\            
292 & L65  &  4.30 $\pm$   0.37&  1.67&  0.43$\pm$   0.05&            &   member \\        
294 & L66  &  3.90 $\pm$   0.18&  0.80&  0.48$\pm$   0.04&            &   member \\       
295 &L121 &  3.78 $\pm$   0.10&  0.54&  0.58$\pm$   0.05&             &  member \\   
302 &L122 &  4.35 $\pm$   0.11&  0.75&  0.65$\pm$   0.04&             &  member \\  
321 &L143 &  2.31 $\pm$   0.29&  0.69&  0.42$\pm$   0.08&             &     nm          \\     
327 &L144 &  3.05 $\pm$   0.21&  1.55&  0.56$\pm$   0.08&             &     nm          \\  
330 &L141 &  4.29 $\pm$   0.13&  1.94&  0.57$\pm$   0.04&             &  member  \\ 
331 &L123 &  3.30 $\pm$   0.18&  1.30&  0.66$\pm$   0.11&             &  nm  \\  
333 &L129 &  5.32 $\pm$   0.40&  1.88&  0.62$\pm$   0.13&             &  nm          \\     
334 &L138 &  4.64 $\pm$   0.16&  1.31&  0.52$\pm$   0.05&             &  member  \\ 
336 &L142 &  3.73 $\pm$   0.53&  1.75&  0.71$\pm$   0.19&             &     nm         \\     
339* &L124 &  3.92 $\pm$   0.20&  1.08&  0.55$\pm$   0.06&            &  nm \\  
342 &L126 &  3.08 $\pm$   0.07&  0.79&  0.63$\pm$   0.04& nm(M; FM)    &  nm         \\  
347 &L125 &  2.12 $\pm$   0.13&  1.39&  0.60$\pm$   0.09&  nm(M; FM)   &  nm   \\

\noalign{\smallskip}
\hline
\end{tabular}


\begin{list}{}{}
\item  {$\mathit a: $} ~Identifications from Haug (\cite{H78})
\item  {$\mathit b: $} ~Identifications from Lindoff (\cite{L68}); RG:Red giant, BS:Blue straggler
\item {* :} Membership according to photometric studies
\item {$\mathit c: $} ~$ {\sigma_{1}}^{2} = \sum {(r_{\lambda}/{\epsilon_
{p_\lambda}})^2}/(m-2)$ ; where 
\item {~~}~~~  $m$ is the number of colors and
\item {~~}~~~  $r_{\lambda} = P_{\lambda} - P_{\rm
max}~\exp(-K~{ln}^2({\lambda_{\rm max}/{\lambda}))}$
\item {$\mathit d: $} ~Membership according to Mermilliod et al. (M) (\cite{MMU08}) and/or Frinchaboy
\& Majewski (FM) (\cite{FM08})
\end{list}

\end{table}

\clearpage

\begin{figure*}

        \includegraphics[width=13cm]{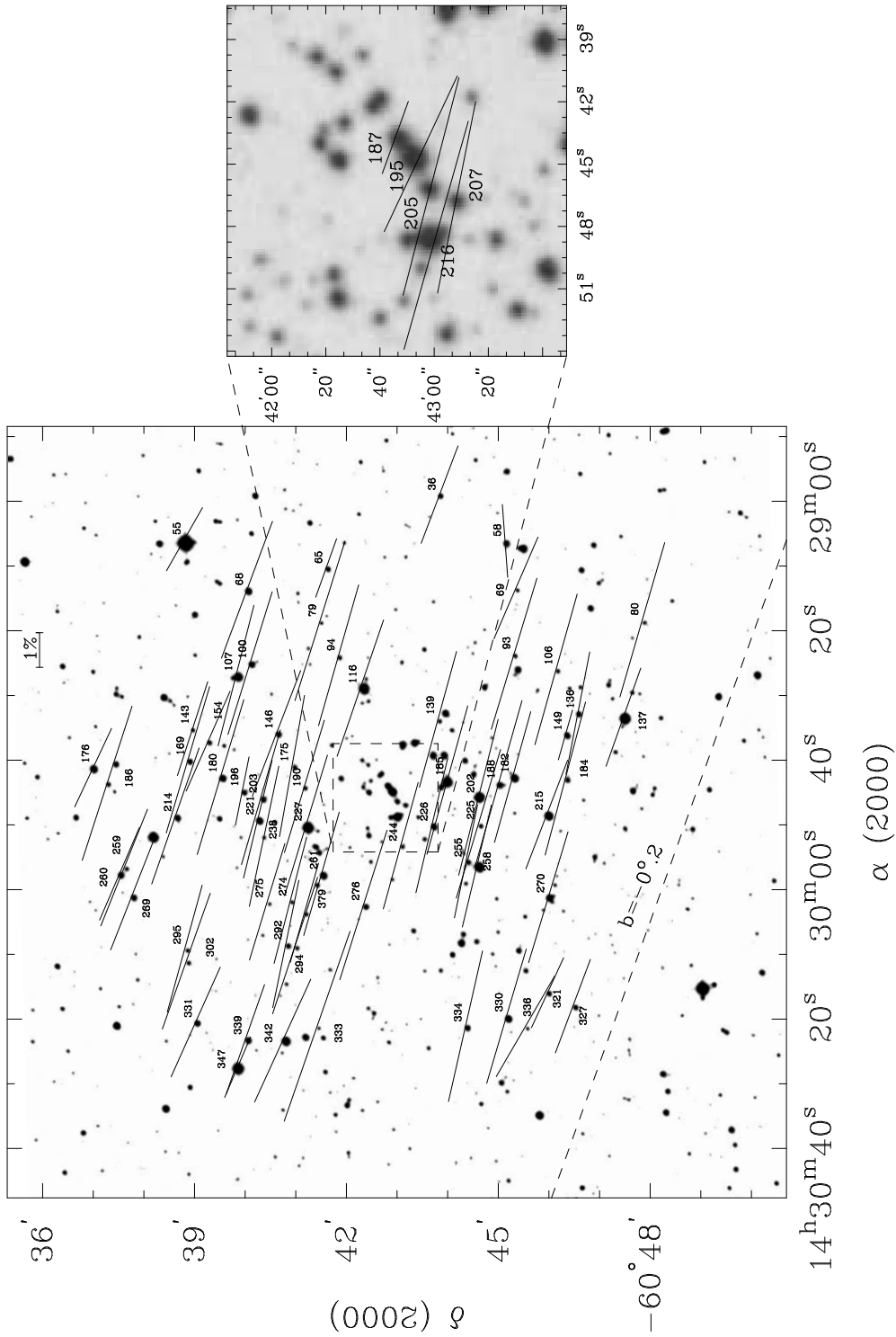}

        \caption{Projection on the sky of the polarization vectors (Johnson V filter)
of the stars observed in the region of NGC 5617. The dot-dashed line is the
Galactic parallel  $b=-0^{\circ}.2$. The length of each vector is proportional to the polarization percentage. The background image of the cluster is a DSS image from the Space Telescope Institute}
\end{figure*}

\begin{figure*}
        \includegraphics[width=13cm,angle=-90]{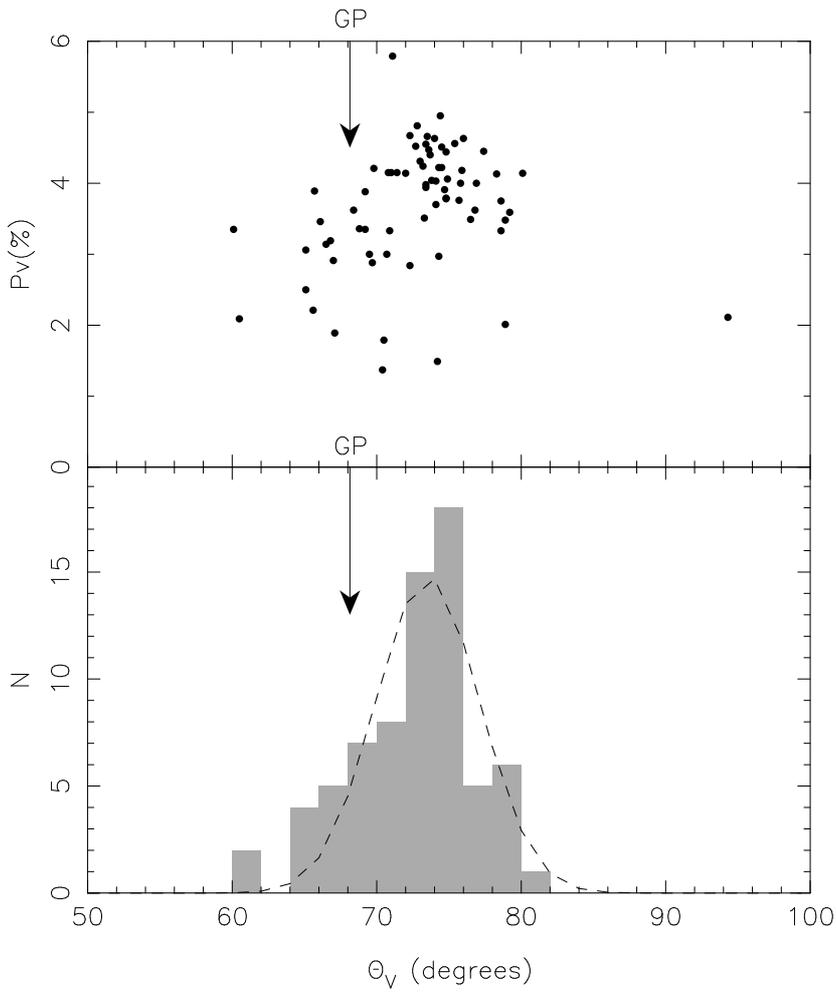}
        \caption{Upper plot: V band polarization percentage of the stellar flux
$P_{V}(\%)$ vs. the polarization angle $ \theta_{V}$ for each star.
Lower plot: Histogram of the polarization angle ($ \theta_V$) for the stars observed in the region. The arrow
shows the location of the projection of the Galactic plane in the region.}
\end{figure*}

\begin{figure*}
         \includegraphics[width=16cm]{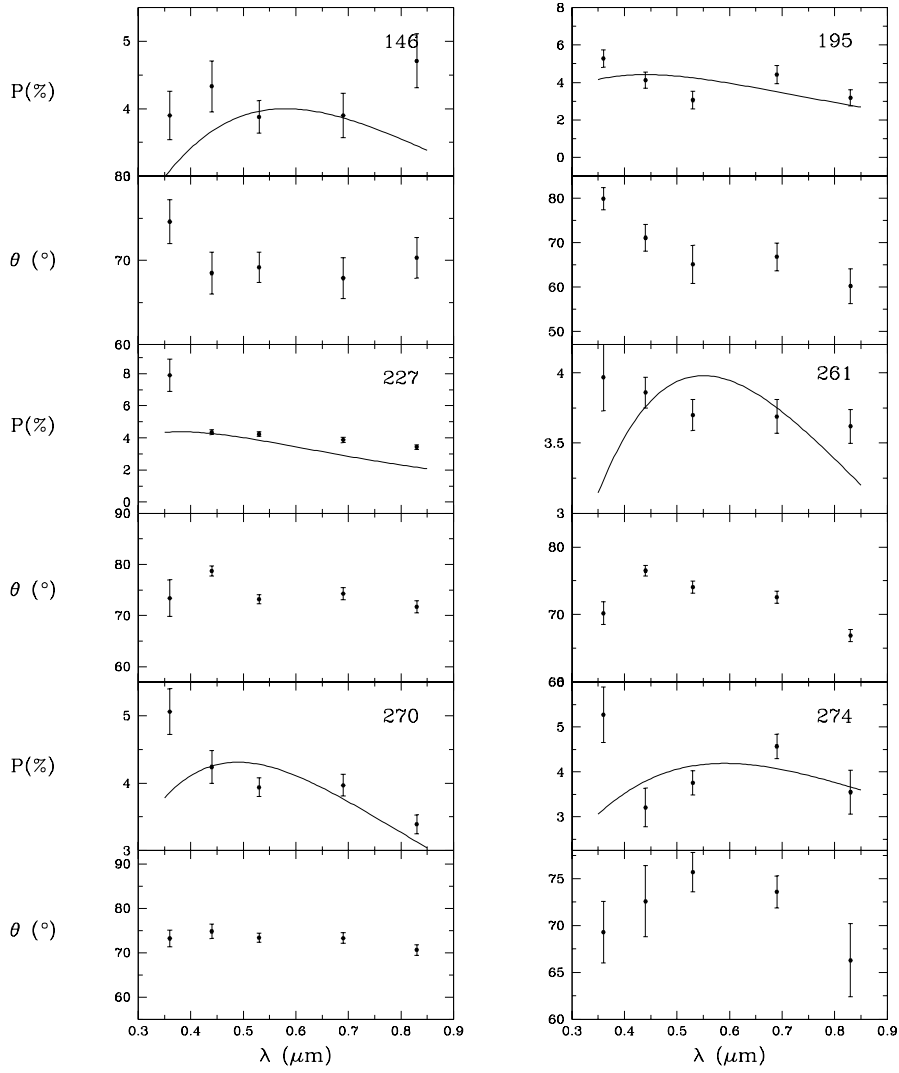}
         \caption{The plot displays both the polarization and position angle dependence on wavelength, for some of the stars with
indications of intrinsic polarization. The numbers in the polarization dependence plots are the star ID., and immediately below
is the position angle dependence for each star.}
\end{figure*}

\begin{figure*}
        \centering
        \includegraphics[width=13cm]{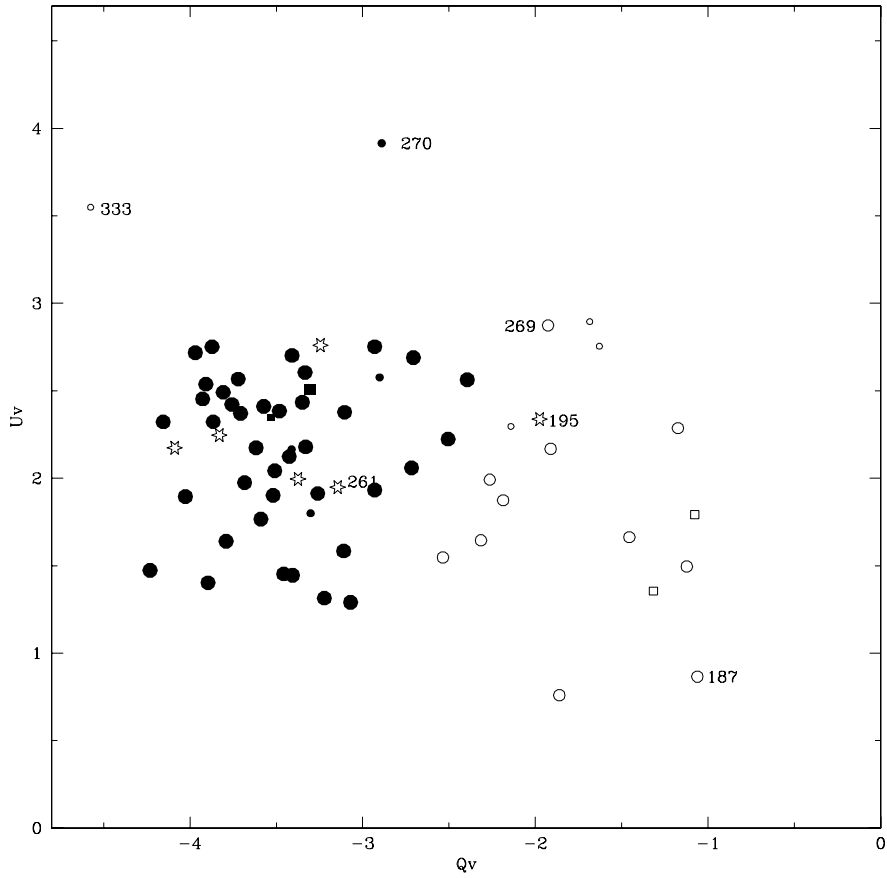}
        \caption{Q- and U- Stokes parameters for the V bandpass. Filled and open circles are for members and non-members,
respectively. Starred points represent
blue straggler stars, and squares red giants. Small symbols are used for stars with intrinsic polarization
according to Table 2.}

\end{figure*}

\begin{figure*}
        \centering
        \includegraphics[width=13cm]{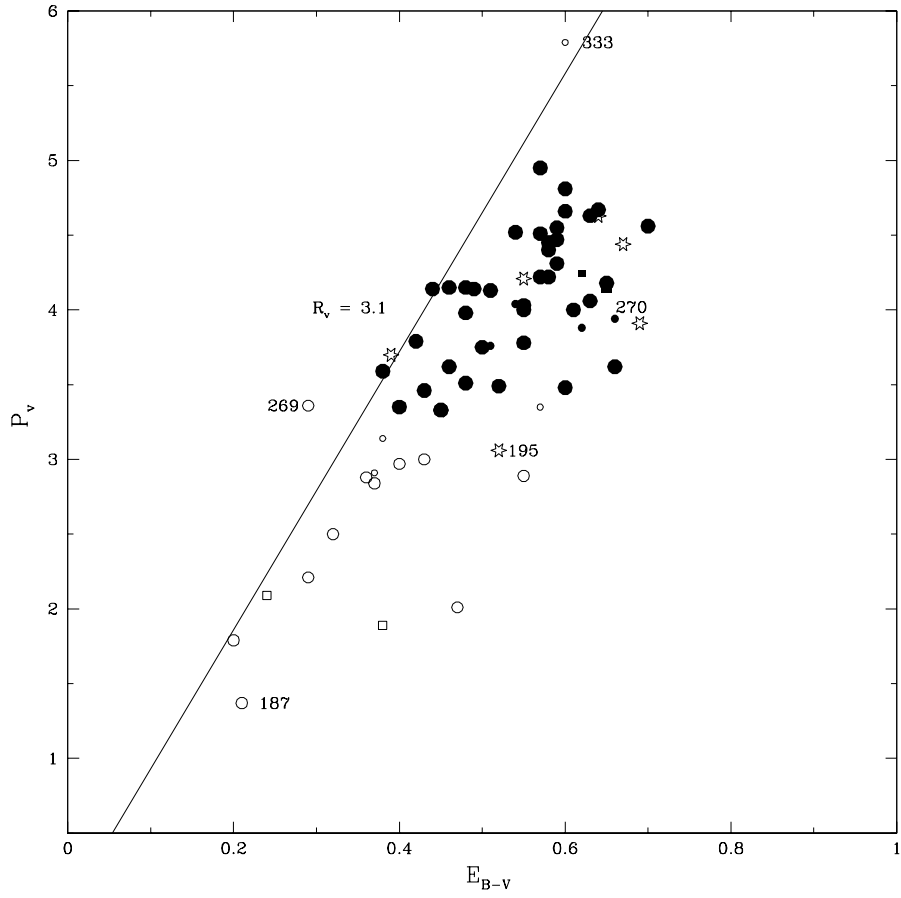}
        \caption{Polarization efficiency diagram.
The line of maximum efficiency is drawn adopting $ R_{v}$ = 3.1. Symbols are the same as in Fig. 4.}     
\end{figure*}

\end{document}